\documentclass[aps,twocolumn,showpacs]{revtex4}
\usepackage{graphicx,amsmath}

\newcommand{\eq}[1]{Eq. (\ref{#1})}
\newcommand{\fig}[1]{Fig.~\ref{#1}}

\newcommand{\ver}{ {\bf r} }
\newcommand{\vom}{\hat{\omega}}

\begin{document}

\title{Long-time self-diffusion of Brownian Gaussian-core particles}

\author{H. H. Wensink, H. L{\"{o}}wen, M. Rex, C. N. Likos and S. van Teeffelen}
\affiliation{Institut f\"ur Theoretische Physik II: Weiche Materie,
Heinrich-Heine-Universit\"at D\"usseldorf, Universit\"atsstra\ss e 1, D-40225 D\"usseldorf, Germany}

\date{\today}

\begin{abstract}
Using extensive Brownian dynamics computer simulations, the 
long-time self-diffusion coefficient
is calculated for Gaussian-core particles as a function of the number density. Both spherical and rod-like particles interacting via  Gaussian segments are considered.
For increasing concentration we find that the translational self-diffusion behaves non-monotonically
reflecting the structural reentrance effect in the equilibrium phase diagram. Both in the limits of zero and infinite 
concentration, it approaches its short-time value.
The microscopic Medina-Noyola theory qualitatively accounts for the translational long-time diffusion.
The long-time  orientational diffusion coefficient for Gaussian rods, on the other hand,
remains very close to its short-time counterpart for any density. Some
implications of the weak translation-rotation coupling for ultrasoft
rods are discussed.
\end{abstract}

\pacs{66.10.Cb; 61.20.Ja; 82.70.Dd} 

\maketitle

\section{Introduction}

Particles interacting via penetrable pair potentials exhibit fascinating new clustering and
reentrance effects \cite{likoscpc,Cluster,Gauss2,Mladek,Frenkelscience} which are absent for diverging potentials such as hard spheres and 
inverse-power potentials. 
A well-studied model for a penetrable interaction 
is a Gaussian potential \cite{Gauss2,Gauss0,Gauss1,Gauss3,Gauss4} which mimics the effective 
interactions between two polymer coils in a good solvent \cite{Louis} and applies also
to dendrimer solutions \cite{dendrimers1,dendrimers2,dendrimers3}.
This potential can  be generalized towards  a Gaussian-segment model for rod-like particles
in order to describe bottlebrush polymers with a stiff backbone \cite{rexwensink}, see also \cite{Saija}. 
In the Gaussian-core system,
two particles  pay a finite  energy penalty if they are sitting on top of each other. If they
overlap completely there is no repulsive force any longer.

In the present paper we focus on equilibrium dynamical correlations  of Gaussian 
Brownian fluids. In particular the long-time self-diffusion coefficient is simulated
as a function of the particle density. Recently the dynamical behavior 
of spherical Gaussian particles has been explored by
molecular dynamics studies \cite{Mausbach} which is suitable for polymer melts but
neglects the hydrodynamic friction of a  solvent.
Here, we consider solutions of colloidal or polymeric particles and therefore overdamped
Brownian dynamics 
is appropriate where the friction of the solvent is included. We further study
the long-time translational and orientational self-diffusion of Gaussian segment rods
in the isotropic phase as a function of rod concentration.

As a result, we find that the long-time self-diffusion coefficient behaves non-monotonically with
density, both for spheres and rods. 
For zero density (i.e. the single particle limit) the long-time self-diffusion
coefficient is clearly identical to the short-time diffusion constant which is entirely dominated 
by solvent friction. What is less obvious is that for high densities with multiple
overlap of particles the self-diffusion again tends to its short-time counterpart.
In fact in the limit of very high densities, a Gaussian particle feels many neighbors
around a distance where the Gaussian potential has its inflection point and these give rise
to a diverging number of interaction kicks. Therefore one could have expected a higher
diffusion coefficient than the short-time value.
However, we show that correlations between the neighboring particles enforce a normal diffusive
behavior of an effective ideal gas in this limit.

Between these two extreme limits, for finite densities, the long-time self-diffusion
coefficient is smaller than its  short-time counterpart. The minimal
value is roughly at the point of maximal fluid structure. 
A similar non-monotonic behavior has been found for molecular dynamics \cite{Mausbach}
where the ballistic limit of zero-density leads, of course, to a diverging long-time 
self-diffusion coefficient.
 Furthermore we find
that the long-time orientational self-diffusion coefficient practically coincides
with its short-time behavior. The latter fact implies that there is no significant
translation-rotation coupling in the Gaussian segment model for penetrable rods.

We compare our simulation data with the microscopic theory of 
Medina-Noyola \cite{Medina-Noyola,LowenSzamel} which
relates the long-time self-diffusion coefficient to the fluid pair correlation
and find qualitative agreement for spheres. The same holds for the 
translational diffusion of rods if the theory of Medina-Noyola \cite{Medina-Noyola} is applied 
to the translational degrees of freedom alone.

The paper is organized as follows: in Section II we describe in detail
the procedure of the Brownian dynamics computer simulation. Results for the long-time self diffusion
are presented and discussed in Section III. Finally Section IV is devoted 
to more general 
remarks and conclusions.

\section{Brownian dynamics computer simulations}

The Brownian dynamics (BD) simulations are based on a
finite-difference integration of the overdamped Langevin equations
for $N$ interacting anisometric particles in three dimensions \cite{Loewen1,Loewen2}. The trajectory of each
particle $i$ is characterized by its position $\ver _i(t)$ and orientation $\vom
_i(t)$ at time $t$. If hydrodynamic interactions are neglected, the
update equation for the position of particle $i$ can be written in the
following way 
\begin{equation}
\ver_{i}(t+ \Delta t) = \ver _{i}(t)  + \frac{\Delta t}{k_{B}T}{\bf
  D}_{0}^{T} \cdot {\bf F}_{i}(t) + \Delta \ver_{i} +
\mathcal{O}  \{ (\Delta t)^2 \}, \label{rupdate}
\end{equation}
with $k_{B}T$ the thermal energy and  ${\bf F}_{i}(t)$  the total force acting on the
center-of-mass. The latter is derived from the pair potential
 which will be specified later. Furthermore, ${\bf D}_0^{T}$ represents the
 short-time diffusion tensor which in case of uniaxially symmetric particles
 (e.g. cylinders) can be cast into the form
\begin{equation}
{\bf  D}_{0}^{T} = D_{0}^{\parallel} ( \vom _i \otimes  \vom _i ) +
D_{0}^{\perp} \left ( {\bf \hat{I} - } \vom _i \otimes  \vom _i \right
), \label{diften}
\end{equation}
in terms of the translational diffusion coefficients parallel ($D_{0}^{\parallel}$) and
perpendicular ($D_{0}^{\perp}$) to the particle axis, with ${\bf
  \hat{I}}$ the unit tensor and $ \otimes $ a dyadic product. 

The contribution $\Delta \ver_{i}$ denotes a
random displacement of the particle due to collisions with the solvent
molecules. Similar to \eq{diften}, it is
convenient to decompose it into contributions parallel and perpendicular to the
particle axis. Introducing two orthogonal unit vectors, $\hat{e}_{1i}$
and $\hat{e}_{2i}$, perpendicular to $\vom _{i}$  we can express the noise term
in \eq{rupdate}  as
\begin{equation}
\Delta \ver_{i} =  \Delta r
  ^{\parallel}  
\vom_{i}(t) +  \Delta r ^{\perp} _{(1)}  
\hat{e}_{1i}(t) +   \Delta r  ^{\perp } _{(2)}   \hat{e}_{2i}(t).
\end{equation}
Here,  $ \Delta r ^{\parallel} $ and  $  \Delta r
^{\perp } _{(1,2)}   $  represent
Gaussian random displacements parallel and perpendicular to the symmetry
axis.  Both stochastic quantities have zero mean and 
their variance is $2 D_{0}^{\parallel} \Delta t $ and  $2 D_{0}^{\perp} \Delta
t $, respectively.

The orientational update equation for $\vom_{i}(t)$ reads
\begin{equation}
\vom_{i}(t+ \Delta t) = \vom _{i}(t)  + \frac{\Delta t}{k_{B}T}
  D_{0}^{R}  {\bf T}_{i}(t) \times \vom_{i}(t) + \Delta \vom_{i} +
\mathcal{O}  \{ (\Delta t)^2 \}. \label{oupdate}
\end{equation}
Here, $D_{0}^{R}$ denotes the short-time rotational diffusion
coefficient and ${\bf T}_{i}(t)$  the total center-of-mass torque acting
on particle $i$. The noise contribution,
\begin{equation}
\Delta \vom_{i} =  x_{1} \hat{e}_{1i}(t) + x_{2} \hat{e}_{2i}(t),
\end{equation}
is generated by means of two uncorrelated random Gaussian numbers,
$ x_{1} $ and  $x_{2}$,
both  with zero mean and variance $2 D_{0}^{R} \Delta t$. After each
step the new orientations $\vom _{i}(t + \Delta t)$ have to be renormalized
to ensure that $|\vom _{i}| =1$ at all times.

Obviously, for spherical particles the translational
Brownian motion is completely decoupled from the orientations and the
translational diffusion tensor \eq{diften} becomes diagonal, i.e.  ${\bf D}_{0}^{T} =
D_{0}^{T} {\bf \hat{I}}$. In this case, we need only consider the  update equation for
the positional coordinates \eq{rupdate}. 
All update equations are exact up to order
$\mathcal{O}(\Delta t)$ which suffices for the present purpose,
provided $\Delta t$ is chosen small enough.
 For a detailed discussion  of a second order
update algorithm the reader is referred to Ref. \cite{Loewen2}.

The pair potential of the particles is given by an ultrasoft
Gaussian potential. For spherically symmetric
particles we have:
\begin{equation}
v_{2}(r) = \epsilon \exp \left [ - ( r/ \sigma ) ^2 \right ],
\end{equation}
where $\sigma$ is the potential range which will henceforth serve as our unit
of length. The amplitude is fixed at $\epsilon = 5 k_{B}T $. According to \cite{Gauss2}, the associated
reduced temperature $T^\ast = k_{B}T/\epsilon = 0.2 $ is much higher than
the upper freezing temperature  $T^{\ast} \cong 0.01 $ which guarantees a stable
fluid state at any density.

\begin{figure}
\includegraphics[clip=,width= 0.4\columnwidth]{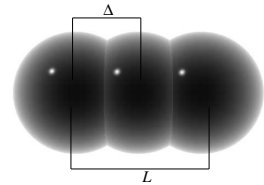}
\caption{\label{segment} Soft rod of length $L$ composed of $N_{S}=3$ segments
  with intersegment distance $\Delta$.}
\end{figure}

Apart from ultrasoft spheres we will also consider systems of {\em
  Gaussian rods} \cite{rexwensink}. Each rod has length $L$ and is composed of $N_{S}$
spherical segments placed at equidistant positions along the rod main
axis with segment-segment distance $\Delta = L/(N_{S}-1)$.   The interaction potential between
two segments from different rods is again a Gaussian. The pair
potential between two rods $i$ and $j$ is then given by a sum over all
segment interactions:
\begin{equation}
  \label{eq:gauss}
    v_{2}(\mathbf{r}_{i},\mathbf{r}_{j} ;\hat{\mathbf{\omega}}_{i},\hat{\mathbf{\omega}}_{j}) =
    \epsilon
    \sum_{\alpha = -K}^{K}  \sum_{\beta = -K}^{K}   \exp [
    - ( |\mathbf{r}_{\alpha \beta}| / \sigma )^2 ],
\end{equation}
where $K=(N_{\mathrm{S}}-1)/2$ and  $\mathbf{r}_{\alpha \beta}=(\mathbf{r}_{i}+
\alpha \Delta \hat{\mathbf{\omega}}_{i})-(\mathbf{r}_{j}
+\beta \Delta \hat{\mathbf{\omega}}_{j})$ the distance between segment $\alpha$ on rod
$i$ and $\beta$ on rod $j$ ($i\neq j$). We will consider slightly anisometric rods with $N_{\mathrm{S}}=3$ segments
and $L=2\sigma$ (see \fig{segment} for a sketch). Furthermore the
segment-segment potential amplitude is $\epsilon = 5 k_{B}T$.
The short-time
diffusion coefficients of the rods depend on one-particle
hydrodynamic effects. For these, we take the analytical results
obtained for hard ellipsoids of length $L$ and aspect-ratio $p > 1$
reported by Tirado and co-workers \cite{tirado}:
\begin{eqnarray}
  D^{\perp}_{0}&=&\frac{3 D_{0}^{T}}{2p}(\ln p +0.839+0.185/p + 0.233/p^{2}),\\ 
  D^{\parallel}_{0}&=&\frac{3 D_{0}^{T}}{p}(\ln p -0.207+0.980/p -  0.133/p^{2}),\\ 
  D^{R}_{0}&=&\frac{18}{\pi p^3}\frac{D_{0}^{T}}{\sigma ^2}(\ln p -0.662+0.917/p - 0.050/p^{2}),
\end{eqnarray}
with $D_{0}^{T} = k_{B}T/6\pi \eta_{s} \sigma$ the short-time diffusion constant of
a sphere with radius $\sigma$ and $\eta_{s}$ the shear viscosity of
the solvent. In the above, we have implicitly identified $L=p\sigma$, with $p$ the
{\it hydrodynamic} aspect-ratio of the Gaussian rods. In order to enforce a substantial translation-rotation coupling 
 we take a value $p=5$ which is larger than the {\it interaction} aspect-ratio $L/\sigma = 2$.

A natural unit of time is the Brownian time $\tau_{B} =
\sigma^2/D_{0}^{T}$ defined as the typical time a Gaussian particle needs
 to diffuse over a distance comparable to its own dimension. 
Let us further introduce $\bar{D}_{0}^T = || \oint d \vom {\bf D}_{i}^{T} || $, the isotropic
orientational average of the diffusion tensor \eq{diften}. For the
spheres $\bar{D}_{0}^{T} = D_{0}^{T}$ while for the rods
\begin{equation}
\bar{D}_{0}^{T} = \frac{1}{3} D_{0}^{\parallel} + \frac{2}{3} D_{0}^{\perp}.
\end{equation}
With this result, we may compute the ratio of the single-rod rotational and
translational relaxation times, i.e. $\tau_{0}^{R}/\tau_{0}^{T} =
\bar{D}_{0}^{T}/p^2 \sigma^2 D_{0}^{R} \cong 0.264$, showing that
the short-time orientational dynamics is much faster than the
translational. The quantity  $\bar{D}_{0}^{T}$ also provides the natural scale for the long-time
translational self-diffusion coefficient $D_{L}^{T}$, defined as
\begin{equation}
D_{L}^{T} = \lim _{t \rightarrow \infty} \frac{1}{6t} \frac{1}{N} \left \langle
   \sum_{i=1}^{N} \left( \ver_{i} (t) - \ver_{i}(0) \right
  )^2 \right \rangle , 
\end{equation}
where $\langle \cdots \rangle$ is a canonical average. An alternative
definition is provided  by a differential expression
\begin{equation}
D_{L}^{T} = \lim _{t \rightarrow \infty} \frac{1}{6} \frac{d}{dt} \frac{1}{N} \left \langle
   \sum_{i=1}^{N} \left( \ver_{i} (t) - \ver_{i}(0) \right
  )^2 \right \rangle . \label{tdif}
\end{equation}
Both expressions should in principle yield identical results in the long-time
limit. In practice however, they will differ slightly  and the difference
can be used to assess the error in $D_{L}^{T}$.

Following Ref. \cite{Loewen2} we can define the long-time {\it
  rotational} diffusion coefficient $D_{L}^{R}$ as follows
\begin{equation}
D_{L}^{R} = - \lim _{t \rightarrow \infty}   
\frac{W_{n}(t)}{t}, \label{drot}
\end{equation}
where $W_{n}(t)$ is an orientational correlation function measuring
the mean-square displacement on the unit sphere. It is given by
\begin{equation}
W_{n}(t) =  \frac{1}{n(n+1)} \ln \left \langle
   \mathcal{P}_{n} ( \vom (t) \cdot \vom
   (0)  ) \right \rangle , \label{wn}
\end{equation}
with $\mathcal{P}_{n}$ a Legendre polynomial. Similar to
\eq{tdif}, we may also employ the differential analogue. If the 
rotational motion is a  diffusion process on the unit
sphere, the dynamics is captured by the Debye diffusion equation which
predicts $D_{L}^{R}$ to be {\it independent} of $n$ \cite{Loewen2}.

In our simulations we used a cubic simulation box of volume $V$ with
periodic boundary conditions in all three directions. The number of
particles depends on the number density. If we choose to cut off all
segment-segment pair
interaction for which $v_2 / \epsilon < v^{\text{cut}}_2 / \epsilon$ the length $L_{B}$
of the simulation box must be at least twice the corresponding cutoff
range and so $L_{B}/\sigma > -2 \ln [ v^{\text{cut}}_2 /\epsilon ] $. For a given
density the number of particles must therefore obey  $N  >  (\rho
\sigma ^3) L_{B}^3$, while imposing a minimum of $N=500$  at small
densities. Using $ v^{\text{cut}}_2 / \epsilon = 10^{-6} $, a run at
$\rho \sigma ^3  = 10 $ requires  $N=4000$ and one at $\rho \sigma ^3
= 25$ $N=13000$. The time step must be reasonably small and was fixed at
$0.0005 \tau _{B}$. Initial configurations were generated by putting the particles at
random positions. For the rod systems, a
parallel nematic initial configuration was adopted. After a long equilibration
period of at least $10 \tau_{B}$ statistics were gathered (during a
period of about $15 \tau_{B}$) and the time-dependent correlations  were monitored
during an interval of about $5 \tau_{B}$. The latter turned out to be
sufficiently large to reach the long-time limit. 

\section{Results for the long-time self-diffusion coefficients}

\begin{figure}
\includegraphics[clip=,width=\columnwidth]{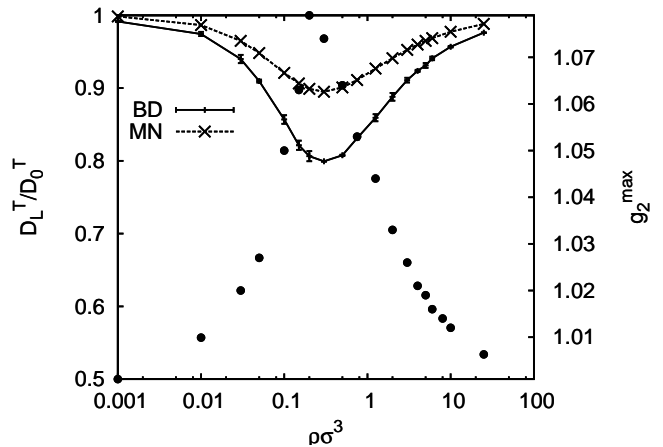}
\caption{\label{spheres} Long-time translational self-diffusion coefficient divided
  by its low-density limit $D_{L}^{T}/D_{0}^{T}$ versus the density
  $\rho   \sigma ^3$ (on a log scale) for Gaussian
  spheres from BD simulations and the theory of Medina-Noyola (MN). On the
  right vertical axis, filled dots give  the amplitude
  $g^{\text{max}}_{2}$ of the first  maximum of the equilibrium pair correlation
  function $g_{2}(r)$.  }
\end{figure}
 Results for  the long-time diffusion coefficient for spheres are
 shown in \fig{spheres}. A clear non-monotonic behavior is observed.
 Both for very small and very high densities, the diffusivity comes
 very close to that of a single particle. At the highest density
 simulated ($\rho \sigma ^3 = 25$) the long-time diffusion constant has regained  about
 $98$ \%  of its short-time value indicating that the diffusion has
become virtually ideal in the high-density limit. The density $\rho
\sigma^3 \cong 0.3$ at which the diffusivity becomes minimal  is in agreement
with the results of Ref. \cite{Mausbach}, and lies close to the density $\rho\sigma^3 \cong 0.23$ for 
which the Gaussian core model displays its `turning point' in the
reentrant melting transition \cite{Gauss2}.

 To compare our data with  microscopic theory we
 have included the prediction from the  Medina-Noyola theory for
 self-diffusion \cite{Medina-Noyola}.The theory comprises  an  analysis of the
 effective Langevin equation of a tagged spherical colloid in a
 medium of interacting neighbor particles.  In the absence of
 hydrodynamic interactions, the following expression for  $
 D_{L}^{T}$ is proposed:
\begin{equation}
D_{L}^T = D_{0}^{T} \left (  1 + \frac{\rho}{6} \int d \ver \left [
    g_{2}(r) - 1 \right ]^2 \right )^{-1},
\end{equation}
where the only input is the static pair correlation function
$g_{2}(r)$ which is obtained from the simulation. Although the theory is certainly not
reliable from a quantitative point of view, as we observe in \fig{spheres}, the non-monotonic behavior
is clearly recovered. This suggests that there is a qualitative
correspondence between the long-time diffusive behavior and the static
correlations as embedded in $g_{2}(r)$. This  
 becomes more explicit when we compare the diffusion data with the
 maximum amplitude in the pair correlation function, also shown in
 \fig{spheres}.
\begin{figure}
\includegraphics[clip=,width= \columnwidth]{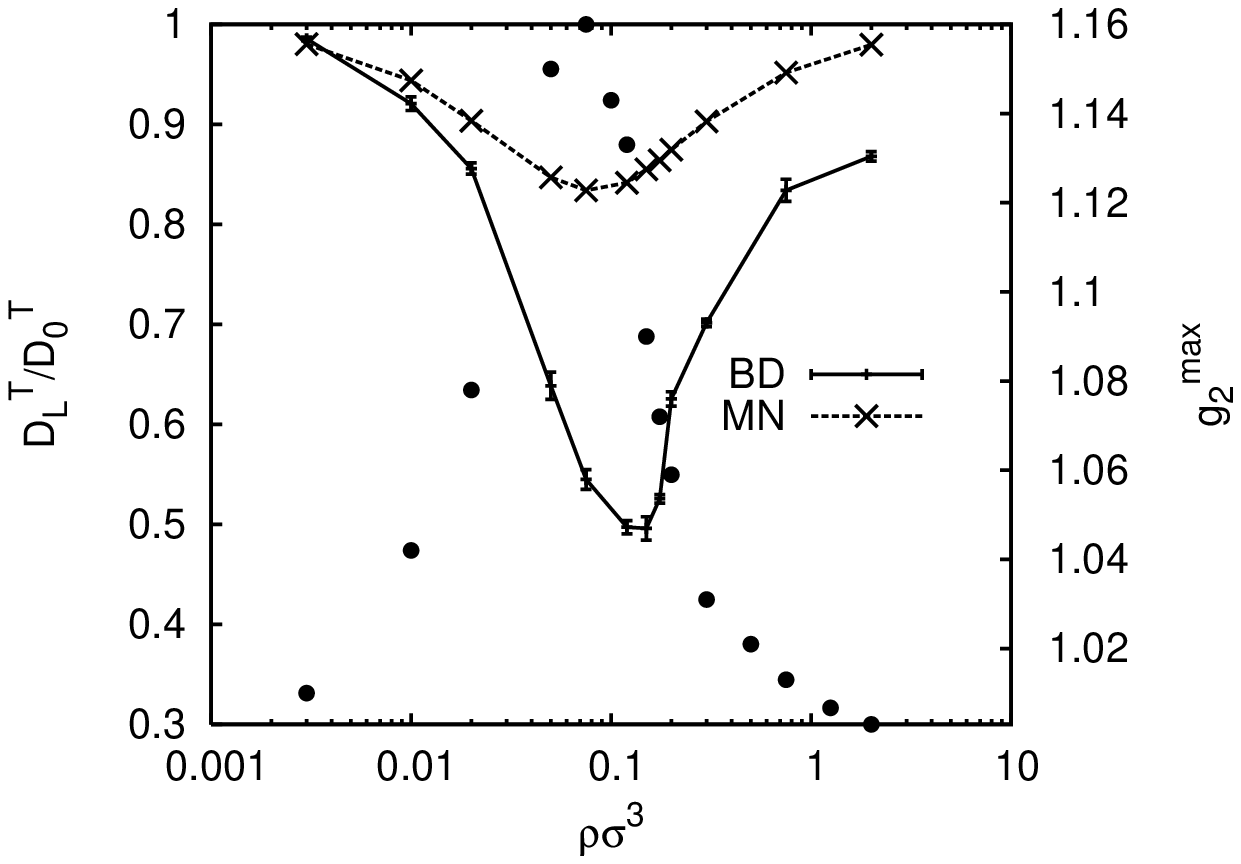}
\caption{\label{rods} Same as \fig{spheres} for the Gaussian segment rods.}
\end{figure}
The behavior for the rods is qualitatively the same as for spheres, see
\fig{rods}. Also here the translational diffusion constant varies non-monotonically
with density and approaches the short-time limits at small  and large
densities. The only notable difference is that the fluid structure is
somewhat more  pronounced here. As a result, the normalized
diffusion constant reaches a minimum value that is smaller than that for spheres.
Again, the theory of Medina-Noyola now taken with the center-of-mass
pair correlations as an input overestimates the simulation data but
shows the correct trend.

\begin{figure}
\includegraphics[clip=,width=\columnwidth]{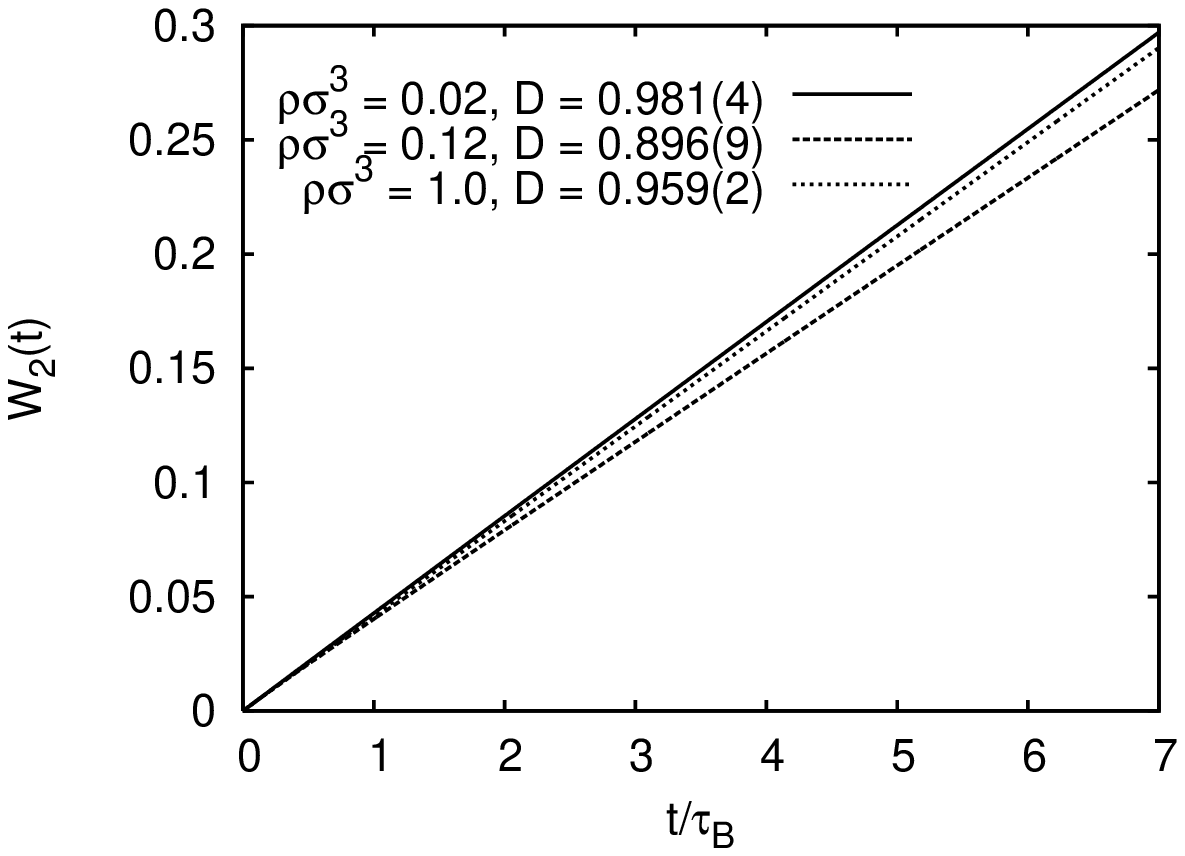}
\includegraphics[clip=,width=\columnwidth]{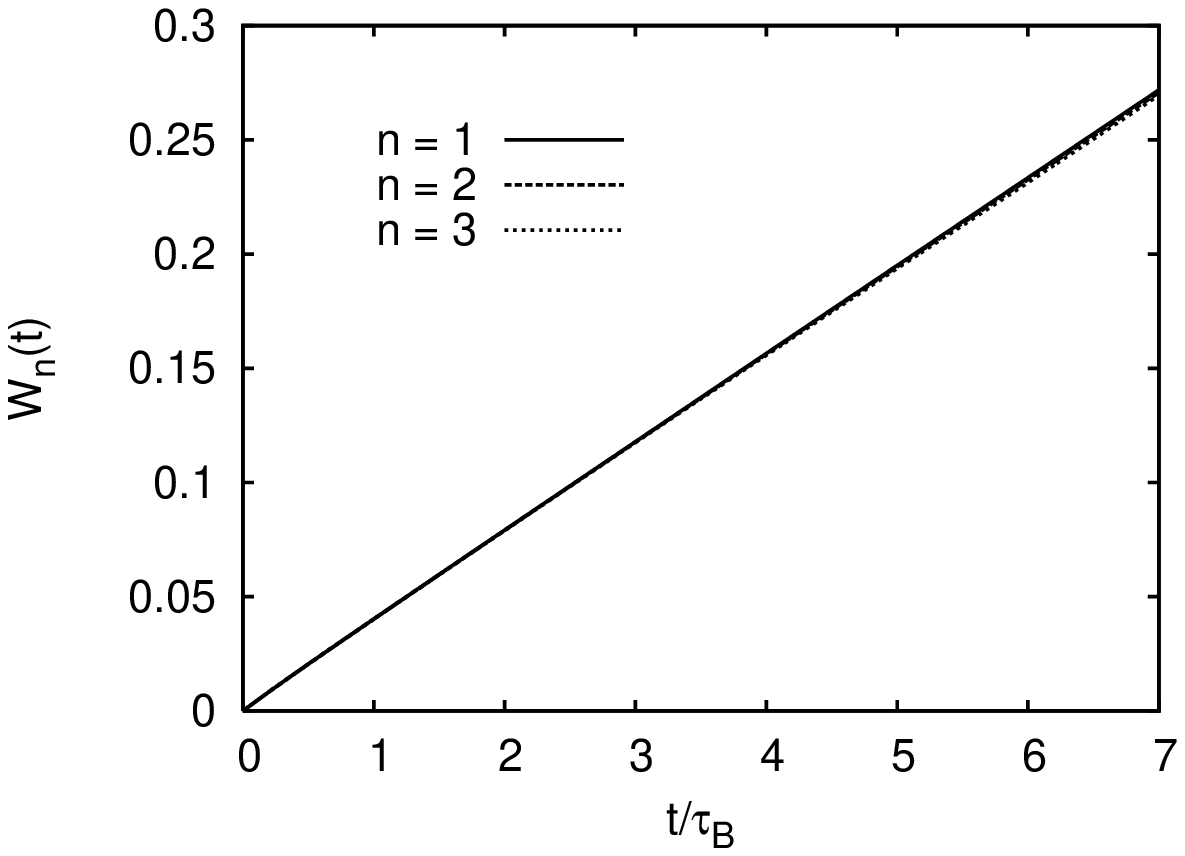}
\caption{\label{rota} (top) Mean-square displacement on the unit
  spheres $W_{2}(t)$ [see \eq{wn} ] for Gaussian rods at various
  densities. The corresponding long-time rotational diffusion coefficients
  $D=D_{L}^{R}/D_{0}^{R}$ are indicated where the number in brackets
  gives the error of the last digit. (bottom) $W_{n}(t)$ for $n=1,2,3$
  for the system with density $\rho \sigma ^3 = 0.12$.}
\end{figure}

Contrary to  $D_{L}^{T}$, the rotational counterpart in \fig{rota} seems 
to be weakly affected by the density. At the point of maximum fluid
structure ($\rho \sigma ^3 \cong 0.12$) the long-time rotational diffusivity
has  dropped to only about 90 \% of the maximum i.e. short-time
value. Moreover, an investigation of $W_{n}(t)$ at minimum diffusivity
shows that the mean-square orientational displacement does not
depend on $n$. From this we may conclude that the rotational relaxation on
the unit sphere is purely diffusive for long times. This behavior is not found in isotropic systems
with unbounded rod potentials such as hard spherocylinders or Yukawa
segment models \cite{Loewen2}.  The distinct discrepancy between
the density-dependence of the translational and rotational diffusivity suggests
that the coupling between  orientational and translational degrees of
freedom is very small for the systems of ultrasoft rods considered here.

Some insight as to the status of the translation-rotation coupling can be gained by considering the {\it effective} interaction of a rod
pair in a spatially homogeneous fluid:
\begin{equation}
v_{2} (\vom_{i}, \vom_{j}) = \int d  \ver _{ij}  v_{2}(  \ver _{ij} ;
\hat{\mathbf{\omega}}_{i},\hat{\mathbf{\omega}}_{j}) \label{v2},
\end{equation}
with $\ver _{ij} = \ver_{j} - \ver_{i}$. Inserting \eq{eq:gauss} and some
algebra leads to  $ v_{2}( \vom _{i}, \vom_{j} ) =
\text{const}$. This result holds for {\it any} bounded
segment-segment potential and implies that, irrespective of the rod aspect-ratio, all static orientational
correlations are rendered zero by the random-phase approximation
for the excess free  energy \cite{Cluster,Gauss2}. For a spatially uniform rod fluid the  latter is simply proportional to a double orientational 
average of
$v_{2}(\vom_{i}, \vom_{j})$. This will give the same outcome for any normalized orientational
distribution. Since the ideal free energy of a nematic fluid is
always higher than that of the isotropic, the possibility of a stable
nematic state is fully excluded. Of course the above argument does not rule out a possible freezing transition
occurring within an isotropic rod fluid. In fact, recent investigations for
other soft rods such as parallel Gaussian-core particles \cite{Saija} and Yukawa rods \cite{wensinkyuk} seem to point to a pronounced stability of
columnar liquid-crystalline order in these systems. Finally, we remark
that for soft rods with large aspect-ratios  a phase transition
from an isotropic toward a nematic fluid {\it may} be possible at low densities where rod
correlations are much better described by the Onsager functional
\cite{onsager}  than the random-phase approximation.  

\section{Conclusions}

In conclusion, we have simulated the long-time self-diffusion
in concentrated Brownian  systems of rod-like and spherical particles which
interact via a Gaussian core  and are thus penetrable.
As reflected by the statics, the system is getting ideal in the high-density
limit where the random-phase approximation for the fluid 
structure becomes asymptotically exact.
We think that the trends are independent of details in the interaction
potentials provided that clustering \cite{Cluster} is avoided.

We finish with a few remarks.
First of all, one should consider the hydrodynamic interactions mediated by the solvent.
These are neglected in the Brownian dynamics simulations, but can be treated using 
more sophisticated (and time-consuming) schemes like lattice-Boltzmann, Stokesian
or the stochastic rotation dynamics \cite{durlofsky,Coveney,Tanaka,Padding,Yeomans}.
Second, there is a need to derive microscopic models for the Brownian motion of stiff rods
with soft interactions on the basis of the Smoluchowski equation involving  mode-coupling approximations.
These approaches then would go beyond a simple effective Medina-Noyola
theory in treating explicitly the 
orientational degrees of freedom. Our simulation results 
may provide benchmark data to test these theories. As to the
statics, it would be worthwhile to map out the freezing behavior of
Gaussian segment rods in the regime of high density and low
temperature. Finally, the existence of a stable nematic phase for Gaussian rods
with large aspect-ratio remains an open question.

\acknowledgments

Financial support from the Deutsche 
Forschungsgemeinschaft under  SFB-TR6 is gratefully acknowledged.

\bibstyle{revtex}

\bibliography{felix}

\end{document}